\documentstyle[12pt]{article}
\def\ii{\'{\char'20}}

\begin{document}

\begin{titlepage}
\title{ \begin{flushright}
{\normalsize UB-ECM-PF 93/17}
\end{flushright}
\vspace{1cm}
\large \bf Manifestations of Space-Time Multidimensionality in
Scattering of Scalar Particles }

\author{\large\bf Demichev A.P.
  \thanks{E-mail address: demichev@compnet.msu.su}  \\
Nuclear Physics Institute, Moscow State University, \\
119899, Moscow, Russia  \\
\and
 \large \bf Kubyshin Yu.A.
\thanks{On sabbatical leave from Nuclear Physics Institute, Moscow State
University, 119899 Moscow, Russia}
\thanks{E-mail address: kubyshin@ebubecm1.bitnet} \\
Departament d'Estructura i Constituents de la Mat\'{e}ria \\
Universitat de Barcelona \\
Av. Diagonal 647, 08028 Barcelona, Spain
      \and
 \large \bf P\'{e}rez Cadenas J.I. \\
Nuclear Physics Institute, Moscow State University,\\
Moscow 119899, Russia }

\date{30 August 1993}

\maketitle

\begin{abstract}

We analyze a possibility of experimental detection of the contribution
of the Kaluza-Klein tower of heavy particles to scattering
cross-section in a six-dimensional scalar model with two dimensions being
compactified to the torus with the radii $R$. It is shown that there is
a noticeable effect even for the energies of colliding particles
below $R^{-1}$ which may be observed in future collider experiments
if $R^{-1}$ is of the order of $1 TeV$.

\end{abstract}

\end{titlepage}

\section{Introduction}

Many of theories beyond the standard model exploit two important
hypotheses: space-time supersymmetry and multidimensionality (see for example
\cite{GSW}). Unfortunately, the complete realization of these
hypotheses  in  field  theoretical  or   string   models   meets
considerable
problems. The most important problem to be solved in SUSY models is
to find a mechanism providing SUSY-breaking at the scale of about
$1 \  TeV$. Similar
problem of finding a mechanism of spontaneous compactification at the
energy scale of a few $TeV$ exists in models with extra dimensions.
Also, it is well known that the supersymmetry
in principle lowers under compactification of a part of
space-time dimensions \cite{DNP}. Thus, it is quite natural to suppose
that the SUSY-breaking scale $M_{SUSY} \sim 1\div 10$ Tev is close to
the compactification scale $M_C$. Another argument backing this
assumption arises from possible cancellations of non-logarithmic
corrections to coupling constant renormalizations, this was discussed in ref.
\cite{Ant}. With such compactification scale one can expect that
an evidence of the existence of extra dimensions will be seen in future
experiments at supercolliders.

Another serious problem in multidimensional field theoretical calculations
is non-renormalizability of the theory. Perhaps the true complete description
of fundamental interactions is given by some
ultraviolate finite theory. Nowadays there is a common belief that such
theory does exist and this is the superstring theory [1].
Unfortunately, calculations in the framework of the superstring theory
are very complicated and model dependent. To study low energy effects
one has to work within an effective multidimensional non-renormalizable
model. In such model physical amplitudes can be calculated by performing
renormalizations order by order of perturbation expansion. For this
one has to add to the Lagrangian counterterms with higher derivatives,
they are supposed to emerge from the complete (superstring) theory as a
consequence of the existence of superheavy particles of Plank mass
\cite{GSW}. Whereas in the framework of a complete
finite quantum theory these counterterms
of the low energy sector could be calculated explicitly, in the
effective non-re\-nor\-mali\-zable
model, we are discussing here, necessary counterterms are added by hand
and coupling constants of the new vertices, given by these counterterms,
are considered as phenomenological ones. Fortunately, the contributions
of these counterterms to the finite part of the amplitudes are of the order
$(s/\kappa ^2)^n$ [1], where $\sqrt{s}$ is the energy of
colliding particles, $\kappa $ can be re\-garded as the cha\-rac\-teristic
scale of the complete theory($\kappa \sim M_{Pl}$ in the case of
super\-strings) and $n \geq 1$. Thus, when $\sqrt{s} \leq M_C \ll\kappa $
these contributions can be neglected.

In the present paper we estimate possible manifestations of effects due
to the presence of extra dimensions in a scalar $\Phi ^{4}$-model
on six-dimensional space-time $M=M^{4} \times T^{2}$, where $M^{4}$ is the
four-dimensional Minkowski space-time and $T^{2}$ is the two-dimensional
torus. We believe that this simple model captures many features of more
realistic multidimensional models, in particular qualitative behaviour
of scattering amplitudes calculated here we expect to be rather
generic. We would like to mention
also that there are some phenomenologically consistent Kaluza-Klein type
models just in six dimensions (see e.g. \cite{RDSS}). Similar
calculations were done for a toy scalar field
model in (2+2)-dimensional space-time $M=M^{2} \times T^{2}$ in ref.
\cite{DIKT}.

A few more remarks are in order. The question of physical effects in
multidimensional theories but from a different point of view was addressed in
\cite{KS}. There the authors used the results of the high energy
experiments to get the upper bound on the size $R$ of the space of extra
dimensions assuming that the first heavy Kaluza-Klein mode had not been
yet observed directly. Our approach in the present paper is different in
two aspects. First, the authors of \cite{KS} considered only one
(the first) heavy Kaluza-Klein mode as an exited vector boson. In our
calculations here we take into account the whole Kaluza-Klein
tower of particles and, hence, all peculiarities
of calculations in non-renormalizable theories. Second,
we assume that $R^{-1}\sim M_{SUSY}\sim 1\div 10\ TeV$ and look for
possible experimental evidence of heavy Kaluza-Klein modes.

In the next section we describe our model and renormalization
condition. In sect.3 the results of 1-loop calculations for
amplitudes and scattering cross section are presented. Sect.4 is
devoted to the analysis of these results. Sect.5 contains
some conclusions.

\section{The model and renormalization}

We consider here a model of one real scalar field $\Phi (x,y)$ on
the six-dimensional space-time $M^{4} \times T^{2}$ with the radii
of the torus being equal to $R = M_{C}^{-1}$. The action is
given by
\begin{eqnarray}
S=\int d^4 xd^2 y\{ \frac{1}{2} g^{MN} \Phi (x,y) \partial _M
\partial _N \Phi (x,y)- \frac{m^2}{2}\Phi ^2(x,y)- & &\nonumber \\
\label{1}\\
\frac{g_6}{4!}\Phi ^4(x,y)\ - \frac{h_6}{4!}\Phi ^2(x,y)\Box \Phi ^2(x,y)
+ \ldots \} & & \; ,\nonumber
\end{eqnarray}
where $M,N=0,...,5$; the metric $g_{MN}={\rm diag}(-1,1,...,1)$,
$g_6$, $h_6$ are six-dimensional bare coupling constants, dots in the
integrand denote possible higher derivative terms. The last term is
necessary for the renormalization of the 4-point Green function at one
loop order. The
field $\Phi (x,y)$ is periodic in coordinates $y^1,\; y^2$:
\[
\Phi (x,y+2\pi R)=\Phi (x,y); \qquad R=M_C^{-1},
\]
and can be expanded in Fourier series:
\begin{eqnarray*}
& &\Phi (x,y)= \frac{1}{(2 \pi R)^{2}} \sum_{n_1,n_2=-\infty}^{\infty}
 \varphi _{\vec n}
(x)\, {\rm exp}\{i\vec y\vec n/R\}\, ,\nonumber \\
& &\vec n=(n_1,n_2)\, ,\quad \vec y=(y^1,y^2)\, ,\nonumber \\
& &\varphi _{-\vec n}=\varphi _{\vec n}^* \; .\nonumber
\end{eqnarray*}
Substituting the Fourier expansion into the action (1) we get
\begin{eqnarray}
S & = &\int d^{4}x \left[ \frac{1}{2} \sum_{\vec n}\varphi _{-\vec n}
(\partial^2-M_{n}^2)\varphi _{\vec n}
+\frac{g_{B}}{4!} \sum_{\vec n,\vec k, \vec p}\varphi _{\vec n}\varphi
_{\vec k}\varphi _{\vec p}\varphi _{-(\vec n+\vec k+\vec p)} +
 \right.  \label{2} \\
 & + & \left. \frac{h_{B}}{4!} \sum_{\vec n,\vec k, \vec p}\varphi _{\vec n}
 \varphi_{\vec k} \Box_{(4)} \varphi _{\vec p}
 \varphi _{-(\vec n+\vec k+\vec p)} -
 \frac{h_{B}}{4!} \sum_{\vec n,\vec k, \vec p}\frac{(\vec n, \vec k)}
 {R^{2}} \varphi _{\vec n} \varphi_{\vec k} \varphi _{\vec p}
 \varphi _{-(\vec n+\vec k+\vec p)} + \ldots \right], \nonumber
\end{eqnarray}
where $(\vec n, \vec k) = n_{1}k_{1}+n_{2}k_{2}$ is the scalar product,
\begin{equation}
 M_{n}^2=m^2+\frac{n^2}{R^2}=m^2+M_{C}^{2} n^2 \ , \ \ \
  n^{2} = n_{1}^{2} + n_{2}^{2}   \label{mass}
\end{equation}
and
\[ g_{B} = \frac{g_{6}}{(2 \pi R )^{2}}, \ \ \
   h_{6} = \frac{h_{6}}{(2 \pi R )^{2}}   \]
are four-dimensional bare coupling constants.

The field $\varphi _{0} \equiv \varphi _{\vec 0} (x)$ with the light
mass $m\ll M_C$ in our simple scalar model would be an analog of the
usual vector fields and fermion fields in the Standard Model, whereas
the Kaluza-Klein tower of fields $\{ \varphi _{\vec n}\, ,\;
\vec n \neq 0 \}$ with heavy masses $M_{n}$ would be an analog of
a set of fields of a multidimensional extension of the Standard Model.
The action (\ref{2}) describes the
four-dimensional theory with infinite number of massive fields, and it is
important to notice that, as a consequence of the six-dimensional
nature of the model, interactions between modes are determined by a few
coupling constants only.

As a process imitating the present possibilities of collider
experiments it is reasonable to consider $(2\ light \ particles)
\rightarrow (2\ light\ particles)$ scattering assuming
that $m\ll M_C$ and the scattering energy satisfies
$E^{2} \leq 4 M_{1}^{2}$. Since $h_{B}$ is expected to be proportional
to $\kappa^{-2} \sim M_{Pl}^{-2}$
(see discussion in the Introduction) it is natural to assume that
$h_{B} M_{C}^{2} \sim g_{B}^{2}$. So,
for the process under consideration the interaction term
with the coupling constant $h_{B}$ will be taken into account in the
next-to-leading order only. Then the
scattering amplitude of the light $\varphi _{0}$-particles in
the leading approximation is defined by the usual tree
contribution of the vertex $\sim \varphi _{0}^4 $. One can see that
heavy modes do not contribute at this level. The reason is that
there are no vertices with
only one heavy field in the sum in eq. (\ref{2}) as a consequence of
rotational symmetry on the torus $T^{2}$. This property is rather generic
and holds for other types of scalar interactions
and all compact spaces of extra dimensions. Hence, the lowest order where
we can hope to find some new effects due to the Kaluza-Klein tower of modes
is the one loop order.

To perform 1-loop calculations for scattering of light particles
besides the vertex
\begin{equation}
  \frac{g_{B}}{4!} \, \varphi _0^4     \label{selfint}
\end{equation}
with the light fields only the following relevant vertices must be taken
into account
\begin{equation}
  \frac{g_{B}}{2} \, \varphi _0^2 \sum_{\vec n}\varphi _{\vec n}\varphi _
{-\vec n} + \frac{h_{B}}{4!} \varphi _{0}^{2} \Box _{(4)} \varphi _{0}^{2}
 \; .   \label{interact}
\end{equation}

We will show that due to these vertices the massive Kaluza-\-Klein
fields give con\-si\-de\-rable contribution to the cross section of the
scattering process under in\-ves\-ti\-gation.

Regularized two light particle scattering amplitude to the one-loop order
in our model has the following form (see, for example, \cite{BS}):
\begin{eqnarray}
{\cal A}_{reg} (p_{12}^{2},p_{13}^{2},p_{14}^{2}) & = &
  g_B \left( 1-\frac {g_B} {32\pi ^2}\sum_{\vec n {\bf :}\ \vec n^2 < K^2}
  \left[ I^{reg}_{n}(p_{12}^{2})+
I^{reg}_{n}(p_{13}^{2})+I^{reg}_{n}(p_{14}^{2}) \right] \right) \nonumber \\
      & + & h_{B} \frac{p_{12}^{2}+p_{13}^{2}+p_{14}^{2}}{12}
                                                       \ ,    \label{3}
\end{eqnarray}
where $p_{ij}^{2} = (p_{i}+p_{j})^{2}$, $p_{i}$, $i=1,2,3,4$ are
the external momenta of the 1-loop diagram satisfying the conservation
law $p_{1}+p_{2}+p_{3}+p_{4}=0$ and $I^{reg}_{n}(p_{12}^{2})$,
$I^{reg}_{n}(p_{13}^{2})$, $I^{reg}_{n}(p_{14}^{2})$
are regularized 1-loop contributions of the $\vec n$th mode
given by the usual four-dimensional momentum integral
\begin{equation}
I^{reg}_{n}(p^{2})=\frac{i}{\pi^2}\int ^{\Lambda^2}d^4k
\frac{1}{[(p-k)^2- M_{n}^2+i\varepsilon]
(k^2-M_{n}^2+i\varepsilon)}                   \label{4}
\end{equation}
The last integral is regularized by momentum cut-off, other appropriate
regularizations (e.g. the dimensional regularization) will also work
here. Thus, we have two regulators in the model: the momentum cut-off
$\Lambda$ regularizing the four-dimensional integrals and the cut-off $K$
regularizing the sum over ${\vec n}$.

Let us discuss now the renormalization of our amplitude.
The sum of the 1-loop integrals in eq. (\ref{3}) corresponds to a 1-loop
integral of the type (\ref{4}) in the original six-dimensional theory
(\ref{1}) which is quadratically divergent. Thus, if the calculations were
carried out directly in six dimensions one would have to make
two subtractions to obtain finite result. Since
compactification does not influence the ultraviolate properties of
the theory also two subtractions are needed in general to make the amplitude
(\ref{3}) in four-dimensions to be finite. The additional
divergence, which is due to the six-dimensional nature of the comlete
model, reveals itself in the divergence of the sum in (\ref{3}) after
the logarithmical divergence of each term $I_{n}^{reg}$ is removed.
It is for this reason that the regulator $K$ have been introduced.
Similar to usual quadratic divergencies in four dimensional models
this additional divergence gives rise to quadratic corrections to the
renormalization of the coupling constant thus reducing considerably
the range of validity of the perturbation theory and, because of this,
is quite undesirable. Fortunately, in the case under
consideration the quadratic divergencies vanish in the sum of
three channels in the amplitude (\ref{3}) on the mass shell if
the renormalization is carried out according to standard
condition
\begin{equation}
 {\cal A}_{reg} (\mu_{s}^{2},\mu_{t}^{2},\mu_{u}^{2}) = g  \label{renorm}
\end{equation}
at the physical subtraction point satisfying
\begin{equation}
\mu^2_s+\mu^2_t+\mu^2_u=4m^2 \ .    \label{s.p.}
\end{equation}
Here $g$ is the physical (renormalized) coupling constant.

Indeed, expressing $g_{B}$ in terms of the physical coupling constant
the amplitude (\ref{3}) can be written as
\begin{eqnarray}
{\cal A}_{reg} (p_{12}^{2},p_{13}^{2},p_{14}^{2}) & = &
          g \left[ 1-\frac {g}{32\pi ^2}
         \sum_{\vec n {\bf :}\ \vec n^2 < K^2}
        \left( I^{reg}_{n}(p_{12}^{2})- I^{reg}_{n}(\mu _{s}^{2})
       + I^{reg}_{n}(p_{13}^{2}) - I^{reg}_{n}(\mu_{t}^{2})
                                              \right. \right. \nonumber \\
     & + & \left. \left. I^{reg}_{n}(p_{14}^{2}) - I^{reg}_{n}(\mu_{u}^{2})
     \right) \right] + h_{B} \frac{ p_{12}^{2} + p_{13}^{2} + p_{14}^{2} -
     \mu_{s}^{2} - \mu_{t}^{2} - \mu_{u}^{2}}{12} \ .  \label{ampl1}
\end{eqnarray}
Of course, to renormalize the coupling constant $h_{B}$ additional
condition must be imposed. However, this is not necessary for
our purpose. The differences of the type $I^{reg}_{n}(p_{12}^{2})-
I^{reg}_{n}(\mu _{s}^{2})$ are
finite and do not depend on the regulator $\Lambda$. We will represent
them as
\begin{equation}
 I^{reg}_{n}(p^{2})- I^{reg}_{n}(\mu^{2}) =
 J\left( \frac{p^2}{4M^2_n} \right)- J\left( \frac{\mu^2}{4M^2_n} \right)\ ,
                         \label{diff}
\end{equation}
where $J(z)$ is the finite part of the 1-loop integral \cite{BS}
\[
 J(z) = 2 + \int_{0}^{1} dx \ln \left(1 - 4 z x(1-x) \right) \ ,
\]
which is equal to
\begin{eqnarray}
J(z) & = & -i\pi\sqrt{\frac{z-1}{z}}+2\sqrt{\frac{z-1}{z}}
\ln(\sqrt{z}+\sqrt{z-1})\qquad \mbox{if}\ \ z>1 \ ,    \label{8} \\
J(z) & = & 2\sqrt{\frac{1-z}{z}}
\arctan \sqrt{\frac{z}{1-z}}\qquad \mbox{if}\ \ 0<z\leq 1 \ ,   \label{9} \\
J(z) & = & 2\sqrt{\frac{z-1}{z}} \ln (\sqrt{1-z}+\sqrt{-z})
\qquad \mbox{if}\ \ z\leq 0 \ .                    \label{10}
\end{eqnarray}
It is easy to see that for small $z$ the expressions (\ref{9}) and (\ref{10})
have the following Taylor expansion
\begin{equation}
J(z) = 2 \left( 1 - \frac{z}{3} - \frac{2 z^{2}}{15} \right)
       + {\cal O}(z^{3}) \ .  \label{jexp}
\end{equation}

Let us show now that the scattering amplitude (\ref{ampl1}) is actually
finite and does not depend on the cut-off $K$ on the mass shell.
Taking formulas (\ref{diff}) and (\ref{jexp}) into account one gets
\begin{eqnarray}
{\cal A}_{reg} (p_{12}^{2},p_{13}^{2},p_{14}^{2}) & = &
         g \left\{ 1-\frac {g}{32\pi ^2}
         \sum_{\vec n {\bf :}\ \vec n^2 < K^2}
        \left[ - \frac{2}{3} \frac{p_{12}^{2} + p_{13}^{2} + p_{14}^{2} -
     \mu_{s}^{2} - \mu_{t}^{2} - \mu_{u}^{2}}{4 M _{n}^{2}}
                                              \right. \right. \nonumber \\
   & - & \left. \left.  \frac{4}{15} \frac{p_{12}^{4}+
       p_{13}^{4}+p_{14}^{4}-
       \mu_{s}^{4}-\mu_{t}^{4}-\mu_{u}^{4}}{(4 M_{n}^{2})^{2}}
       + {\cal O} \left(\frac{1}{M_{n}^{6}} \right) \right] \right\}
                                                        \label{ampl2} \\
   & + & h_{B} \frac{ p_{12}^{2} + p_{13}^{2} + p_{14}^{2} -
         \mu_{s}^{2} - \mu_{t}^{2} - \mu_{u}^{2}}{12} \ .  \nonumber
\end{eqnarray}
On the mass shell $p_{12}^{2}=s$, $p_{13}^{2}=t$, $p_{14}^{2}=u$, where
$s$, $t$ and $u$ are the Mandelstam variables satisfying the well known
kinematical relation
$$
s+t+u=4m^2 \ .$$
Together with eq. (\ref{s.p.}) this amounts to vanishing of the first
term in the square brackets and the term proportional to $h_{B}$. The rest
of the one-loop contribution in the expression (\ref{ampl2}) is equal to
the finite sum so that the cut-off $K$ can be sent to infinity. We also
see that the amplitude renormalized by the condition (\ref{renorm}) is
independent of the coupling constant $h_{B}$ on the mass shell. Finally
we have
\begin{eqnarray}
{\cal A}_{mass\  shell} (s,t) & = & g \left\{ 1-\frac {g}{32\pi ^2}
 \sum_{\vec n}^{\infty}
        \left[ J \left( \frac{s}{4M_{n}^{2}} \right) -
               J \left( \frac{\mu_{s}^{2}}{4M_{n}^{2}} \right)
             +  J \left( \frac{t}{4M_{n}^{2}} \right)
             -  J \left( \frac{\mu_{t}^{2}}{4M_{n}^{2}} \right)
                                              \right. \right. \nonumber \\
     & + & \left. \left. J \left( \frac{u}{4M_{n}^{2}} \right)
             -  J \left( \frac{\mu_{u}^{2}}{4M_{n}^{2}} \right)
	       \right] \right\}             \ ,  \label{ampl3}
\end{eqnarray}
where $u = 4 m^{2} -s-t$ and the subtraction points are related by eq.
(\ref{s.p.}).

\section{Heavy modes contribution to the total cross section}

In this section we will study the contribution of heavy Kaluza-Klein
particles to the total cross section of the scattering of two light
particles (zero modes). The most interesting range of energies is above
the threshold of the light particle and below the threshold of the first
heavy mode: $4m^{2} \leq s \leq 4(m^{2} + M_{C}^{2})$. Indeed, due to
decoupling of heavy modes at low energies
(see \cite{KCS}) the contribution of the Kaluza-Klein tower below $s=4m^{2}$
is likely to be negligible. On the other hand, if $M_{C} \sim M_{SUSY}$
it is very unlikely that the energies of colliders will exceed the
threshold of the first heavy particle in the near future.
For the purpose of our analysis we found convenient to
introduce the following quantity
\begin{equation}
\Delta ^{(N,0)}(s)=  16 \pi^{2} \frac{\sigma^{(N)}(s)-
\sigma^{(0)}(s)}{g \sigma^{(0)}(s)} \ .    \label{deltadef}
\end{equation}
Here $\sigma^{(N)}(s)$ is the total cross section of the
$(2\ light\ particles) \rightarrow  (2\ light\ particles)$ scattering
calculated using the amplitude of the type (\ref{ampl3}) but with
summation over ${\vec n}$ satisfying $0 \leq |{\vec n}| \leq N$ only.
Physically
this means that only a finite number of heavy particles of the
Kaluza-Klein tower with $|{\vec n}| \leq N$ contribute to
the 1-loop correction to the total cross section $\sigma ^{(N)} (s)$.
The cross section calculated in the complete six-dimensional
theory is equal to $\sigma ^{(\infty)} (s)$. Notice that to compute
the leading order of $\Delta^{(N,0)}(s)$ in $g$ only tree approximation for
$\sigma^{(0)}(s)$ is used in the denominator of (\ref{deltadef}).

We calculate the total cross sections using formulas (\ref{9}), (\ref{10})
and (\ref{ampl3}) and get the following expression for $\Delta^{(N,0)}$
\begin{equation}
\Delta^{(N,0)}(s)= - \sum_{n>0,\vec n^2<N}\left[ J(z_n) - 2 + 2 J(y_{n})
   + \frac{J^{2}(y_{n})}{2 (1-y_{n})}
   - J(\tilde{z}_n) - 2 J(\tilde{y}_n) \right]\ ,
                                                      \label{deltafin}
\end{equation}
where
\begin{eqnarray*}
z_n & = & \frac{s}{4m^2}\xi _{n} \  ; \ \ \  y_{n} = \xi _n - z_n <0 \ ;
                                                      \nonumber   \\
 \tilde{z}_n & = & \frac{\mu_{s}^{2}}{4 m^{2}}\xi_n \ ; \ \ \
           \tilde{y}_{n} = (\xi _{n} -  \tilde{z}_{n})/2  \ ;    \nonumber \\
            \xi _n & = & (1+\frac{M_{C}^{2}}{m^2}\vec n^2)^{-1}\ , \nonumber
\end{eqnarray*}
and we put $\mu_{t}^{2} = \mu_{u}^{2}$. For numerical evaluations
we take
\[ \frac{m^{2}}{M_{C}^{2}} = 10^{-4} \ , \ \ \
   \frac{\mu _{s}^{2}}{m^{2}} = 10^{-2} \ .   \]
Since $\mu_{s}^{2} \ll m^{2} \ll M_{C}^{2}$ the function (\ref{deltadef})
practically depends on the variable
\[ z = \frac{s}{4 M_{C}^{2}} \simeq z_{1} = \frac{s}{4 M_{1}^{2}} \]
only. The plot of $\Delta ^{(\infty)} $ as a function of $z$ is presented
in Fig. 1.

The behaviour of this function shows that the contribution
of the Kaluza-Klein tower of particles increases rapidly when the
energy of colliding particles grows. We see that accummulation of these
contributions is considerable and gives a quite noticeable effect even
for energies much below the threshold of the first heavy mode.
Thus, $\Delta ^{(\infty,0)} \approx 0.76$ for $ s = 0.5 (4 M_{C}^{2})$
and  $\Delta ^{(\infty,0)} \approx 0.17$ for $ s = 0.25 (4 M_{C}^{2})$.
Even for $s = 0.1 (4 M_{C}^{2})$ the effect is not that small:
$\Delta ^{(\infty,0)} \approx 0.03$.

It is clear, of course, that due to the convergence of the sum
in (\ref{deltafin}) the heavier the particle of the Kaluza-Klein
tower the less it contributes to this sum. Hence the curve
$\Delta ^{(\infty,0)}$ actually represents the contributions of a
few first modes. (To draw the plot in Fig. 1 we approximated it
by $\Delta ^{(20,0)}$. We found that this approximation is sufficiently
good, for example, $| \Delta^{(21,0)}(s) - \Delta^{(20,0)}(s) |
< 10^{-9}$ for all $s$ in the range under investigation.)
It is useful to evaluate relative contributions of a few first
heavy modes. In a more realistic model comparison of these kind of estimates
with experimental data would allow us to conclude how many heavy particles
one  actually "detects" in the experiment (see discussion in the next
Section). To analyse this we calculated the function $\Delta ^{(1,0)}$,
representing the contribution of the first heavy mode, for
the same range of energies (Fig. 1) . From the curves in Fig. 1 one can
clearly distinguish the "infinite" number of modes and just one first mode
for the
energies below its threshold. Thus, the difference between them is
about $0.23$ for $s = 0.5 (4M_{C}^{2})$ and about $0.05$ for $s= 0.3
(4M_{C}^{2})$.

To have more illustrative characteristics
let us introduce the quantities
$$
\epsilon_N (z) \equiv\frac{\Delta^{(N,0)}(4zM_{1}^{2})}
               {\Delta^{(\infty,0)}(4zM_{1}^{2})} \ ,  $$
which show the relative contribution of the first $N$ heavy Kaluza-Klein
modes, and the quantities
$$
\delta\epsilon_N (z) \equiv \epsilon _{N}(z) - \epsilon_{N-1}(z)\ , $$
which show the relative contribution of the $N$th mode. The plots for some
$\epsilon _N (s)$ and $\delta\epsilon _N (s)$ for the same range of energies
as before are presented in Fig.2 and Fig.3. Combining the results presented
in Fig. 1 - Fig. 3 we conclude that with accuracy about $5 \div 10 \% $
the function $\Delta ^{(\infty,0)} (s)$ in the energy region
$s\sim\ 0.5M^2\div 0.9M^2$ shows the presence of
at least $3 \div 4$ first heavy modes in the theory.

\section{Discussion and conclusions}

We have shown that effective field theories, obtained from
models in $4+d$ dimensions, provide self-consistent way to calculate
possible effects related to their multidimensional nature.

Though our model is not physical we believe that the results
(for example, rapid growth of $\Delta ^{(\infty,0)} (s)$ above the
threshold of the light particle) capture some general features in the
more realistic theories. In the latter case analogous results could be
{\em in principle} used for comparison with experiments as follows.
We assume that the low-energy sector of the theory, which is the sector
consisting the field $\varphi_{0}$ only, is already well determined. Thus,
the value of the coupling constant $g$ is known (notice, that due to the
multidimensional nature of the complete theory the coupling constant of
the self-interaction of $\varphi_{0}$ and coupling constants of the
interactions of the zero mode with non-zero ones are the same, see
(\ref{2}), (\ref{3}), (\ref{interact})), and the total cross
section $\sigma ^{(0)} (s)$ of the $(2\ light\ particles) \rightarrow
(2\ light\ particles)$ scattering within the low energy sector can
be calculated with sufficient accuracy. Experimentally
we measure the total cross section $\sigma ^{exp}(s)$ and compute
the quantity
\[
\Delta ^{exp}(s)= - 16 \pi^{2} \frac{\sigma^{exp}(s)-
\sigma^{(0)}(s)}{g \sigma^{(0)}(s)} \ .
\]
(cf. (\ref{deltadef})). If above the threshold of the light particle
$\Delta ^{exp} (s) = 0$,
there are no any evidence of heavy Kaluza-Klein modes at given energies.
If $\Delta ^{exp} (s) \neq 0$, there is an evidence of the existence
of heavier particles. Next step would be to see which curve $\Delta ^{(N,0)}
(s)$ fits the experimental data best. If it is the curve with $N=\infty$ or
sufficiently large $N$, then this might be considered as an indirect
evidence of the multidimensional nature of the interactions, at least in
the framework of the given model and given type of compactification. Our
calculations suggest that the effect is rather noticeable even for energies
below the threshold of the first heavy particle (see Fig. 1). Thus,
$\Delta ^{(\infty,0)} \approx 0.03$ for $s = 0.1 (4 M_{C}^{2})$ that, with
our supposition $M_{C} \sim M_{SUSY} \sim (1 \div 10) TeV$ corresponds to
the energy of colliding particles $\sqrt{s} \sim 0.63 (1 \div 10 ) TeV$. Our
numerical estimations show also that a few ($\sim 3\div 4$) first
Kaluza-Klein modes can be "seen" in the range $0.5 < s/(4M_{C}^{2}) <1$
with $5 \div 10 \% $ accuracy.

We would like to mention that we also computed the differential cross
section of the $(2\ light\ particles) \rightarrow (2\ light\ particles)$
scattering. However it does not add essentially new information about the
relative contributions of the heavy Kaluza-Klein particles.

An interesting question is whether different types of compactification
can be distinguished at these energies. The main
difference comes from the structure of the spectrum of the states in
the theory, namely from the multiplicities of states with the same mass and
the rate of the growth of the mass with $n$ (see (\ref{mass})).
This issue is under investigation now. Also it would be interesting
to understand whether one can distinguish between
Kaluza-Klein type models and certain models with infinite tower of composite
particles.

Two more remarks are in order. Here we considered effects when the
energies of the scattering particles are less then the threshold of the
first Kaluza-Klein particle. In this case
momentum behaviour of the effective (running) coupling constant is
mainly defined by the contribution of the lightest mode only \cite{Tay},
\cite{Shir}, \cite{DKL}. Note that renormalization group equation for
the coupling constant $g$, which is physically important at
(relatively low) energies $s\sim M^2_c$,
is independent  of the coupling constant $h$ of the vertices with
derivatives (the latter is important for renormalization only) \cite{KCS}.

Another remark is that there are two cases when heavy modes may give more
considerable contributions. The first case includes multidimensional models
with
non-scalar particles. In such models for certain spaces of extra dimensions
some of the heavy Kaluza-Klein modes may give non-zero contributions
in the tree approximation of perturbation theory. The second possibility is
to consider specific
processes for which the tree approximation is absent and the leading
contribution is given by the 1-loop diagrams even in the zero mode sector
of the theory. Such processes are obviously more sensitive to the heavy
Kaluza-Klein modes. These possibilities are also under investigation.

\section*{Acknowledgements}

The enlightening discussions with E. Boos, F. Cuypers, V.A. Kuzmin,
J.M. Mour\~{a}o and J. Valle are gratefully acknowledged.
This work was partially supported by Russian Higher Education Committee
Grant No.2-63-2-3. Yu. K. acknowledges support from Direcci\'{o}n General de
Investigaci\'{o}n Cient{\ii}fica y T\'{e}cnica (sabbatical grant
SAB 92 0267) during part of this work and thanks the Departament
d'ECM de la Universitat de Barcelona for its warm hospitality.

\newpage

\section*{Figure captions}

\begin{description}
  \item[Fig. 1] Plots of $\Delta^{(\infty,0)}$ and $\Delta^{(1,0)}$
                as functions of $z \equiv s/(4 M_{1}^{2})$.
  \item[Fig. 2] Plots of functions $\epsilon_N (z) = \Delta^{(N,0)}
                (4zM_{1}^{2})/\Delta^{(\infty,0)}(4zM_{1}^{2})$ for
		$N=1,2,3,4,5$; $\epsilon_{\infty} \equiv 1$.
  \item[Fig. 3] Plots of functions $\delta\epsilon_N (z) =
                \epsilon _{N}(z) - \epsilon_{N-1}(z)$ for $N=2,3,4,5$;
		$\delta \epsilon_{\infty} \equiv 0$.
\end{description}

\end{document}